\documentclass{cjaa}                    

\input{epsf.sty}                        
\input{psfig.sty}                       

\begin{document}

   \title{Phenomenological Analogies in TeV Blazars and Black Hole X-ray Binaries in low state}

   \author{You-Hong Zhang
      \inst{}\mailto{}
      }
   \offprints{Y.H. Zhang}                   

   \institute{Department of Physics and Tsinghua Center for Astrophysics (THCA), Tsinghua University, Beijing 100084, P.R. China \\
             \email{youhong.zhang@mail.tsinghua.edu.cn}
          }

   \date{Received~~2004~~~~~~~~~~~~~~~ ; accepted~~2004~~~~~~~~~~~~~~ }

   \abstract{Blazars are well known for the violent variability from the radio to gamma-ray wavelengths. In this paper, we present a brief summary of the X-ray variability recently obtained for TeV blazars. We also point out the probable similarities of the X-ray variability characteristics between blazars and microblazars. Such phenomenological analogies could be interpreted and unified with the synchrotron X-ray emissions from the jets of the two classes of objects that differ in black hole mass by factors of $\ga 10^7$. The phenomenological analogies in blazars and microblazars (both are jet-dominated) might be parallel to those in black hole X-ray binaries in high/soft state and Seyfert galaxies (both are disk-dominated).       
}
   \authorrunning{Y.H. Zhang} 

\titlerunning{Phenomenological Analogies in Blazars and Microblazars}
  
\maketitle

\section{Introduction}

Blazars are an extreme class of Active Galactic Nuclei (AGNs), they show rapid variability on different timescales from years to minutes across the whole electromagnetic spectrum (Urry \& Padovani 1995; Ulrich, Maraschi \& Urry 1997). The emission from blazars is generally believed to be produced by relativistic electrons tangled with the magnetic field in a relativistic jet roughly aligned with our line of sight. TeV blazars are blazars whose emission was detected up to TeV energies. Up to now, there are 6 confirmed TeV blazars, including the three classical blazars Mrk~421, Mrk~501, and PKS~2155$-$304. The overall spectral energy distributions of TeV blazars showed that the synchrotron emission from these sources peaks at high energy (UV/soft X-ray) band. 
This indicates that the X-ray emission from TeV blazars is the high energy tail of the synchrotron emission component, in which the most rapid variability is theoretically expected. Therefore, TeV blazars are the jet/synchrotron dominated X-ray sources, and have been the important targets of various X-ray telescopes such as ASCA, RXTE, BeppoSAX and XMM-Netwon. These observations revealed valuable but very complex variability patterns from these sources (Pian 2002). 

AGNs and black hole X-ray binaries (BHXRBs) are the accreting systems associated with the supermassive ($\sim 10^6-10^9 M_{\odot}$) and stellar mass ($\sim 10 M_{\odot}$) black holes, respectively. It was proposed that all systems dominated by accretion onto black holes are governed by the same physics (Mirabel \& Rodr\'{\i}guez 1999).  In this context, AGNs and BHXRBs should exhibit analogous phenomena, and the scales of length and time in which the phenomena take place are proportional to the black hole masses. This was demonstrated by the fact that state transitions have been observed in both BHXRBs and AGNs. BHXRBs are known to appear in number of states which are power- or accretion rate-dependent, among them two distinct states are the low/hard state (low flux with a hard power law X-ray spectrum) and high/soft state (high flux with a soft X-ray spectrum). Interestingly, similar transitions have been observed in the quasar 3C 120 (Marscher et al. 2002), but in time scale of years rather than minutes.

The X-ray timing properties in Seyfert galaxies and BHXRBs, which differ in black hole mass by factors of $\ga 10^{5}$, are strikingly similar: (1) both show a tight, linear correlation between flux and rms variability amplitude (Uttley \& M${\rm ^c}$Hardy 2001); (2) dedicated long-term X-ray monitoring with RXTE have revealed that the red noise power spectral density (PSD) of Seyfert galaxies are quite similar to those of BHXRBs, with break timescales scaling linearly with black holes mass (e.g., Uttley, M${\rm ^c}$Hardy \& Papadakis 2002; Markowitz et al. 2003); (3) the variability properties are also photon energy and Fourier frequency dependent: (a) the PSD slope above the high frequency break is flatter with the harder energy band; (b) the variabilities between different energy bands are highly coherent (i.e., well correlated at each Fourier frequency) and show time lags, with variations at lower energies leading ahead those at higher energies; and (c) the magnitude of the lag is Fourier frequency dependent, with the delay increasing at lower frequency. These energy and timescale dependent variability properties have been well studied in BHXRBs (e.g., Miyamoto \& Kitamoto 1989; Nowak et al. 1999), but only measured in two Seyfert galaxies, NGC~7469 and MCG-6-30-15 (Nandra \& Papadakis 2001;  Papadakis, Nandra \& Kazanas 2001;  Vaughan, Fabian, \& Nandra 2003). These variability properties appeared remarkably similar to those seen in the high/soft state of BHXRBs such as Cyg~X-1, suggesting the X-ray emission mechanisms are the same in Seyfert galaxies and BHXRBs, except for the timescales (break timescales and time lags) scaling linearly with black hole mass.
  
Recently, Falcke, K\"{o}ding \& Markoff (2004) proposed a specific scheme to unify accreting black hole systems in terms of the mass and accretion rate. In this unification scheme, low-power accreting black holes, blazars and FR I radio galaxies (the unbeamed parent population of BL Lac objects), are the supermassive counterparts of BHXRBs in the low/hard state. Because BHXRBs in the low/hard state always exhibit relativistic jets associated with a sudden drop of the X-ray fluxes, microquasars are usually used to designate BHXRBs in the low/hard state (Mirabel et al. 1992). When the jet axis is close to the line of sight, the source appears as a ``microblazar'' (Mirabel \& Rodr\'{\i}guez 1999), showing analogous phenomena to blazars. Due to the relatively low statistical probability of small viewing angle, ``microblazars'' have not been definitely identified so far. However, it was proposed that a number of ultraluminous X-ray sources (ULXs) in nearby galaxies are microblazars with relativistically beamed jets pointed toward us (Mirabel \& Rodr\'{\i}guez 1999; K\"{o}ding, Falcke \& Markoff 2002). Black hole spin probably also plays a role in the unification model to explain different sub-classes of the accretion systems (e.g., Urry \& Padovani 1995 for AGN unification model). 

Strong X-ray variability is a common phenomenon for BHXRBs, Seyfert galaxies and blazars (van der Klis 1989; Ulrich, Maraschi, and Urry 1997). The similarities between BHXRBs in the high/soft state and Seyfert galaxies in their X-ray timing properties have been frequently mentioned in the literature, however, the X-ray temporal properties between blazars and microblazars (BHXRBs in the low/hard state) are much less compared. Thanks to relatively a large number of X-ray observations, several TeV-emitting blazars indeed showed some similar features to those of BHXRBs in the low/hard state in their X-ray timing properties. In this paper, we will briefly compare the X-ray variability properties of TeV blazars with those of BHXRBs in the low/hard state. We will focus on two important issues, PSD and time lags.

\section{X-ray Power Spectral Density}

The X-ray PSDs of Mrk~421 and PKS~2155$-$304 obtained from BeppoSAX observations were presented in Zhang et al. (1999; 2002) and Zhang (2002). Due to the short length of the BeppoSAX observations, these PSDs did not reveal break timescales. The useful information from BeppoSAX observations is that the sources showed a featureless power law PSD of slope $\sim$ 2 ($P_{\nu} \sim \nu^{-\alpha}$) over the Fourier frequency range of $\sim 10^{-5}$ to  $\sim 10^{-3}$ Hz. However, the breaks must exist, or the total power would diverge toward low frequencies. Indeed, longer observations by ASCA and RXTE showed preliminary evidence for single break power law PSDs for Mrk~421, Mrk~501, and PKS~2155$-$304 (Kataoka et al. 2001). The power law slopes of the PSDs break from about 1 to about 2 from low to high frequency, and the break timescales located at about 1 day (i.e., $\sim 10^5$~s). If TeV blazars are the supermassive counterparts of BHXRBs in the low/hard state, this feature should correspond to the high frequency break of BHXRBs such as Cyg~X-1 in the low/hard state, where the mean high frequency break timescale is $\sim 0.3$~s. The black hole masses of TeV blazars estimated with the stellar velocity dispersion ranges between $10^8-10^9 M_{\odot}$ (Falomo et al. 2003), thus the ratio of black hole mass between TeV blazars and  Cyg~X-1 ($\sim 10 M_{\odot}$) is $\sim 10^7-10^8$. The linear scaling law means that the high frequency break timescale of TeV blazars should be $\sim 3\times 10^6-10^7$~s, roughly consistent with the preliminary ``high frequency break timescales''($\sim 10^6$~s) of TeV blazars after correcting for the relativistic beaming (Doppler factor is about 10). Therefore, preliminary results support the linear scaling law between TeV blazars and BHXRBs in the low/hard state. However, given the large uncertainties in determining the PSD shapes and the break timescales of TeV blazars, in particular we have not seen the low frequency break in a blazar, at the present time it is still premature to unify blazars and microblazars in terms of the X-ray PSDs, and to apply the linearly scaling of break frequency with black hole mass between TeV blazars and microblazars. 

In context of AGNs, thanks to RXTE, well sampled long-term observations of several Seyfert galaxies gave rise to the overall X-ray PSDs mostly similar to the BHXRBs such as Cyg~X-1 in the high/soft state (Uttley, M${\rm ^c}$Hardy \& Papadakis 2002; Markowitz et al. 2003). A single break power law model is appropriate to describe the PSDs of Seyfert galaxies, and the PSD slopes changed from about 1 to about 2 from low to high frequencies, the feature of BHXRBs in high/soft state. These results are consistent with an approximately linearly scaling of break frequency with black hole mass. However, Ark~564 and NGC~3783 PSDs tended to show two breaks, which is similar to those of BHXRBs in the low/hard state. More observations are needed to clarify this issue.

\section{X-ray Time Lags}

The inter-band time lags between the low and high energy variability have been detected in many observations of Mrk~421, Mrk~501, and PKS~2155$-$304 obtained with ASCA, RXTE, BeppoSAX and XMM-Newton. Here we give a brief summary of this issue: (1) the time lags were different from flare to flare (Tanihata et al. 2001; Zhang et al. 1999; 2002; 2004). Both soft lags (soft photons lag hard ones) and hard lags (hard photons lag soft ones) were detected. The amplitude of the lags ranges from 0 to about 3 hours; (2) The time lags appear to be correlated to the flare durations: the longer the flare duration, the larger the lag (Zhang et al. 2002; Brinkmann et al. 2003); (3) The time lags are probably Fourier frequency-dependent: the lag decreases with increasing Fourier frequency (Zhang 2002); (4) The time lags are photon energy-dependent: the lag increases with increasing gap between the two cross-correlated energy bands (Kataoka et al. 2000; Zhang et al. 2002; Zhang 2002; Ravasio et al. 2004).

\subsection{Fourier Frequency Dependence of Time Lags}

With the cross-spectral technique, we obtained for the first time the Fourier frequency dependence of the hard lags of the X-ray variability between the low and high energy bands of a large flare of Mrk~421 obtained with BeppoSAX (Zhang 2002). This dependence is quite similar to those of BHXRBs such as Cyg~X-1 in the low/hard states obtained with RXTE (e.g., Li, Feng \& Chen 1999; Nowak et al. 1999). Both showed a power law trend of hard lags as a function of Fourier frequency with the form $\tau_{\rm lag} \sim \nu^{-1}$, where  $\tau_{\rm lag}$ is the hard lag, and $\nu$ is Fourier frequency. The difference is the hard lag and the Fourier frequency range. The hard lag is the order of 1000~s and ms for Mrk~421 and Cyg~X-1, respectively, which are roughly linearly scaled to the black hole masses.

\subsection{Photon Energy Dependence of Time Lags}

The time lags in TeV blazars were also photon energy-dependent ( Kataoka et al. 2000; Zhang et al 2002). The hard lag of Mrk 421 in the large flare mentioned above also showed photon energy dependence (Zhang 2002). Interestingly, the hard lags of BHXRBs such as Cyg~X-1 in the low/hard state (Nowak et al. 1999; K\"{o}rding \& Falcke 2004) also showed similar photon energy dependence, which are roughly logarithmic. Note that at the present time such dependence are not in direct comparison because the hard lags in TeV blazars were measured with respect to the highest energy band using the cross-correlation function (CCF) method that produced an average value of soft/hard lags in the studied timescale (or Fourier Frequency) range, while the hard lags in Cyg~X-1 were calculated at a specific Fourier frequency, say 2.5 Hz, measured with respect to the lowest energy band using the cross-spectral method. The data with higher time resolution and count rate from XMM-Newton data is able to make a direct comparison between them.

\section{Discussion}

BHXRBs in the low/hard state and TeV blazars show a number of common features in the X-ray variability. Using the X-ray PSDs and hard time lags, in this paper we presented a couple of evidences for the similar X-ray variability patterns between the two classes of objects with black hole mass differed by factors of $\ga 10^7$.

In the context of unification scheme, the X-ray PSDs of TeV blazars are proposed to be similar to those of BHXRBs in the low/hard state, but this assertion must seriously be tested with long-term observations spanned at least over 4 orders of Fourier frequency range, which is able to test whether the PSDs show doubly-break power law and to test the scaling law of the break frequency with black hole masses. In order to unify the black hole accretion system with all masses, it is also essential to test in the future whether Seyfert galaxies really are scaled up BHXRBs in the high/soft state (disk-dominated), and whether (TeV) blazars really are scaled up BHXRBs in the low/hard state (jet-dominated).

 One TeV blazar Mrk 421 showed evidence for a trend of dependence of X-ray hard lag on both Fourier frequency and photon energy similar to those of BHXRBs such as Cyg X-1 in the low/hard state. It is important to confirm this analogy with more data and more sources.
 
The similarities of the X-ray variability behaviors between TeV blazars and BHXRBs in the low/hard state could be interpreted and unified with the jet-dominated synchrotron X-ray radiation. However, up to now the interpretations are distinct in their own context: 

\begin{itemize}
\item Soft/hard lags detected in TeV blazars are usually interpreted with the cooling/accelerating time scales of relativistic emitting electrons, higher energy electrons to produce higher energy photons will have longer accelerating time but shorter cooling time (e.g, Zhang et al. 2002; Zhang 2002);  
\item Hard lags detected in BHXRBs in the low/hard state are generally interpreted by the comptonization of soft photons from the disks by relativistic electrons from either hot corona or from the jets. More scatters are needed for higher energy photons.
\end{itemize}

However, Markoff et al. (2001) and (K\"{o}ding \& Falcke 2004) have suggested that the hard power law spectrum and hard lags could also be attributed to the synchrotron emission from the jets in BHXRBs. It is therefore probable that the hard lags of BHXRBs in the low/hard state have the same origin as in TeV blazars, i.e., the hard lags are related with the accelerating timescales of relativistic electrons. It is worthing noting that soft lags have not been detected in BHXRBs yet. This is probably caused by the contamination of the 0.1-10 keV X-ray emission from the hot corona and/or the inner disk. Therefore, for microblazars, it is essential to compare the time lags at energy range larger than 10~keV, from which soft lags related to the cooling times of relativistic electrons would be expected.

Probably the low/hard state and high/soft state of BHXRBs are the low and high accretion rate modes, respectively, and so are the TeV blazars and normal (or high luminosity) Seyfert galaxies. The accretion rate is the key paprameter for the dichotomy of observational phenomena in both the stellar mass black hole and supermassive black hole accretion systems, respectively.

\begin{acknowledgements}
We thank the anonymous referee for the constructive suggestions and comments that greatly improved this paper. This work is financially supported by the National Natural Science Foundation of China (NSFC) under project 10473006. This work is also partly supported by the Scientific Research Foundation for the Returned Overseas Chinese Scholars, state Education Ministry, and by the Special Foundation of Chinese Academy of Science under project KJCX2-SW-T08. 
\end{acknowledgements}

\label{lastpage}
\end{document}